\documentclass[twocolumn,pre,showpacs]{revtex4}
\usepackage{amssymb}
\usepackage{amsfonts}
\usepackage{amsmath}
\usepackage{graphicx}

\begin{document}

\title{Fluctuation relations with intermittent non-Gaussian variables}
\author{Adri\'{a}n A. Budini}
\affiliation{Consejo Nacional de Investigaciones Cient\'{\i}ficas y T\'{e}cnicas
(CONICET), Centro At\'{o}mico Bariloche, Avenida E. Bustillo Km 9.5, (8400)
Bariloche, Argentina}
\date{\today}

\begin{abstract}
Non-equilibrium stationary fluctuations may exhibit a special symmetry
called fluctuation relations (FR). Here, we show that this property is
always satisfied by the subtraction of two random and independent variables
related by a thermodynamic-like change of measure. Taking one of them as a
modulated Poisson process, it is demonstrated that intermittence and FR are
compatibles properties that may coexist naturally. Strong non-Gaussian
features characterize the probability distribution and its generating
function. Their associated large deviation functions (LDF) develop a
\textquotedblleft kink\textquotedblright\ at the origin and a plateau regime
respectively. Application of this model in different stationary
nonequilibrium situations is discussed.
\end{abstract}

\maketitle


\section{Introduction}

Intermittency is a phenomenon that appears in a wide class of situations
such as chaotic dynamics \cite{chaotic}, atomic \cite{atomic} and nanoscopic 
\cite{blinkinQD} fluorescent systems, single-molecule reaction dynamics \cite%
{molecule}, biological self-organized models \cite{evolution}, or fluid
turbulence \cite{fluids}, just to name a few. It consists in a random
switching amongst qualitatively different system dynamical regimes. Its
stochastic behavior may be ergodic or not. Complexity, non-Gaussian
statistics, and nonequilibrium dynamics are closely related with its
development. In this last context, (Gallavoti-Cohen) FR \cite{gallavoti}
addresses the distribution of entropy production in far-from-equilibrium
steady states. It relates the probability of observing a certain entropy
production rate to the probability of observing the corresponding entropy
consumption rate \cite{gallavoti,kurchan,spohn,maes,crocks,seifert}.
Theoretical and experimental results confirm its validity in diverse
nonequilibrium systems \cite%
{searles,gaspard,kellay,wang,carberry,TwoPoisson,seifertExper,jammed,foffi,UDOColloidal}%
. Interestingly, the symmetry imposed by FR has also been found valid for
non-thermodynamic variables \cite{soodPituto}.

As FR and intermittence are intrinsically related with nonequilibrium
dynamics, it is natural to ask about the possible coexistence of both
properties. The main goal of this contribution is to give a positive answer
to this issue. We demonstrate that there may exist non-equilibrium steady
states whose (non-Gaussian) fluctuations are intermittent and satisfy the
FR. An explicit construction of a stochastic variable with the required
symmetry defines the basis of the present analysis. It is shown that FR can
always be satisfied by the subtraction of two independent stochastic
variables related by an exponential change of measure. Hence, intermittence
is introduced by choosing one of them as a random modulated Poisson process.
Special attention is paid to the asymptotic regime, where FR can be analyzed
through the large deviation functions (LDF) \cite{gallavoti,kurchan,spohn}
of the probability distribution and its associated generating function. In
the parameter regime where intermittence arises, they develop a kink around
the origin and a plateau regime respectively. These strong non-Gaussian
features are similar to that found in different observables such as the
entropy production rate of a colloidal particle \cite{UDOColloidal} and the
(time-average) velocity of a polar granular rod \cite{soodPituto}. While the
specific analysis of these systems is beyond the scope of this contribution,
the present results suggest that intermittence under the constraint of FR
symmetries may be a central ingredient when studying nonequilibirum states
characterized by non-Gaussian fluctuations.

\section{Fluctuation relations with independent stochastic variables}

Our results rely in the following analysis. Let an arbitrary stochastic
variable $x_{st},$ whose probability distribution $P(x)$\ satisfies a FR
defined as%
\begin{equation}
P(-x)=P(x)\exp [-\zeta x],  \label{simetria}
\end{equation}%
where $\zeta $ is a given real positive constant. This symmetry implies that
negative values are exponentially less probable than positive values. In
terms of the generating function $Z(\lambda )=\left\langle e^{-\lambda
x_{st}}\right\rangle =\int_{-\infty }^{+\infty }dxP(x)e^{-\lambda x},$ $%
Z(0)=1,$ the FR reads%
\begin{equation}
Z(\lambda )=Z(-\lambda +\zeta ).  \label{FR_funcionCaracteristica}
\end{equation}%
In order to satisfy the FR, we write $x_{st}$ as the subtraction of two
statistically independent variables $x_{st}^{\pm },$%
\begin{equation}
x_{st}=x_{st}^{+}-x_{st}^{-}.  \label{resta}
\end{equation}%
Then, $Z(\lambda )=Z_{+}(\lambda )Z_{-}(-\lambda ),$ where $Z_{\pm }(\lambda
)$ are the generating functions of $x_{st}^{\pm },$ and whose distributions
are $P_{\pm }(x).$ The condition (\ref{FR_funcionCaracteristica}) is
satisfied by demanding the relation%
\begin{equation}
Z_{-}(\lambda )=\frac{Z_{+}(\lambda +\zeta )}{Z_{+}(\zeta )},
\label{Zetales}
\end{equation}%
which in turn implies 
\begin{equation}
P_{-}(x)=P_{+}(x)\frac{e^{-\zeta x}}{\left\langle e^{-\zeta x}\right\rangle
_{+}},  \label{measures}
\end{equation}%
where $\left\langle e^{-\zeta x}\right\rangle _{+}=Z_{+}(\zeta ).$ This is
one of the central results of this paper. It says us that the FR symmetry (%
\ref{simetria}) is always satisfied by (\ref{resta}) whenever the change of
measure (\ref{measures}) is imposed. By reading $x$ as an energy and $\zeta $
as an inverse temperature, its thermodynamic like-structure is self-evident.
Similar transformations were introduced in Refs. \cite%
{touchette,garrahan,Counting}. The structure of Eq. (\ref{measures}) can
also be read as a variation of the change of measure used in umbrella
sampling \cite{umbrella}.

Note that in Eq. (\ref{measures}) no conditions are required on $P_{+}(x),$
while $P_{-}(x)$ is determined from\ it. Alternatively, one can choose $%
P_{-}(x)$ as the independent distribution. In fact, Eq. (\ref%
{FR_funcionCaracteristica}) can also be satisfied by taking%
\begin{equation}
Z_{+}(\lambda )=\frac{Z_{-}(\lambda -\zeta )}{Z_{-}(-\zeta )},
\label{ZetalesInversa}
\end{equation}%
delivering the relation%
\begin{equation}
P_{+}(x)=P_{-}(x)\frac{e^{+\zeta x}}{\left\langle e^{+\zeta x}\right\rangle
_{-}},  \label{measuresInversa}
\end{equation}%
where $\left\langle e^{+\zeta x}\right\rangle _{-}=Z_{-}(-\zeta ).$ These
last two expressions follow straightforwardly by inverting Eqs. (\ref%
{Zetales}) and (\ref{measures}). Nevertheless, what is important to realize
is that, in general, both kind of solutions lead to different family of
solutions if one assume the same property for each \textquotedblleft
independent\textquotedblright\ distribution.

The relations (\ref{measures}) and (\ref{measuresInversa}) do not involves
time. Therefore, the proposal Eq. (\ref{resta}) guarantees the fulfillment
of the symmetry (\ref{simetria}) at any time.

\subsection*{Time-average variables}

When studying a given nonequilibrium steady state, the stochastic variable
of interest may be defined through a time average,%
\begin{equation}
\mu _{\tau }(t)\equiv \frac{1}{\tau \left\langle v\right\rangle }%
\int_{t}^{t+\tau }dt^{\prime }v_{st}(t^{\prime })=\frac{x_{st}(t+\tau
)-x_{st}(t)}{\tau \left\langle v\right\rangle },  \label{mumu}
\end{equation}%
where $(d/dt)x_{st}(t)=v_{st}(t).$ This definition makes sense when the
average of the integral contribution grows linearly with time. For example, $%
x_{st}(t)$ may represents entropy production and $v_{st}(t)$ its rate \cite%
{UDOColloidal}, or respectively the position and velocity of a
self-propelled particle \cite{soodPituto}. $\left\langle v\right\rangle $ is
the stationary mean value of $v_{st}(t).$

Usually the regime of interest is the stationary one, $t=\infty .$ Hence, we
write $\mu _{\tau }=\lim_{t\rightarrow \infty }\mu _{\tau }(t).$ For this
stationary variable, instead of Eq. (\ref{simetria}), we write the FR%
\begin{equation}
P(-\mu _{\tau })=P(\mu _{\tau })\exp [-\alpha \mu _{\tau }\tau ].
\label{FTTemporal}
\end{equation}%
This condition can also be satisfied in the present approach because the
statistical properties of $\mu _{\tau }$ follows from those of the
stationary increments of $x_{st}(t),$ that is $x_{\tau }^{\infty }\equiv
\lim_{t\rightarrow \infty }[x_{st}(t+\tau )-x_{st}(t)].$ By writing $x_{\tau
}^{\infty }$ as the subtraction of two independent stochastic variables
related by the change of measures (\ref{measures}) [or (\ref{measuresInversa}%
)], it follows $P(-x_{\tau }^{\infty })=\exp [-\zeta x_{\tau }^{\infty
}]P(x_{\tau }^{\infty }).$ The connection between $x_{\tau }^{\infty }$ and $%
\mu _{\tau }$ implies the probabilities relation $P(\mu _{\tau })d\mu _{\tau
}=P(x_{\tau }^{\infty })dx_{\tau }^{\infty }$ \cite{vanKampen}, which lead
to FR Eq. (\ref{FTTemporal}) with $\alpha =\zeta \left\langle v\right\rangle
.$ Consequently, the proposal based on independent\ stochastic
contributions, as in the previous case [Eq. (\ref{simetria})], allows to
fulfill the symmetry (\ref{FTTemporal}) for any value of $\tau .$

\subsection*{Large deviation functions}

In general, a (rate) variable [Eq. (\ref{mumu})] that characterizes a given
non-equilibrium state only satisfies the relation (\ref{FTTemporal}) in a
long time regime, $\tau \rightarrow \infty .$ In fact, the Gallavoti-Cohen
FR is an asymptotic relation in time \cite{gallavoti,kurchan,spohn}. In this
regime, when the probability distribution adopt the asymptotic structure ($%
\lim_{\tau \rightarrow \infty }$) $P(\mu _{\tau })\approx \exp [-\tau
\varphi (\mu _{\tau })],$ the statistics can be analyzed through a large
deviation theory \cite{touchette}. Consistently, the generating function $%
Z(\lambda )=\left\langle e^{-\lambda \tau \mu _{\tau }}\right\rangle $\
scales in the same way, $Z(\lambda )\approx \exp [-\tau \Theta (\lambda )].$
Both $\varphi (\mu _{\tau })$ and $\Theta (\lambda )$ define the LDF's of
the problem. They completely characterize the asymptotic regime. In terms of 
$\varphi (\mu _{\tau }),$ the FR reads%
\begin{equation}
\frac{1}{\tau }\log \left[ \frac{P(+\mu _{\tau })}{P(-\mu _{\tau })}\right]
\approx -\varphi (+\mu _{\tau })+\varphi (-\mu _{\tau })=\alpha \mu _{\tau }.
\label{phi}
\end{equation}%
Through a sadle-point approximation, both LDF can be related by a
Legendre-Fenchel transformation \cite{touchette}%
\begin{equation}
\varphi (\mu _{\tau })=\max_{\lambda }[\Theta (\lambda )-\lambda \mu _{\tau
}],\ \ \ \ \ \ \Theta (\lambda )=\min_{\mu _{\tau }}[\varphi (\mu _{\tau
})+\lambda \mu _{\tau }].  \label{legendre}
\end{equation}%
These relations and Eq. (\ref{phi}) lead to the equivalent formulation of
the FR symmetry \cite{spohn}%
\begin{equation}
\Theta (\lambda )=\Theta (-\lambda +\alpha ).  \label{teta}
\end{equation}

Conditions (\ref{phi}) and (\ref{teta}) are well known expressions of
Gallavoti-Cohen FR symmetry \cite{gallavoti,kurchan,spohn}. As the
independent variables splitting [Eq. (\ref{resta})] allows to fulfill the FR
(\ref{FTTemporal}) at any time, trivially it can also be utilized in the
long time regime. After explicitly writing $\mu _{\tau }=\mu _{\tau
}^{+}-\mu _{\tau }^{-},$ Eq. (\ref{Zetales}) and the asymptotic structure of 
$Z(\lambda )$ allows us to write%
\begin{equation}
\Theta (\lambda )=\Theta _{+}(\lambda )+\Theta _{+}(-\lambda +\alpha
)-\Theta _{+}(\alpha ),  \label{TetaTotal}
\end{equation}%
where $\Theta _{+}(\lambda )$ is the LDF corresponding to the generating
function of $\mu _{\tau }^{+}.$ By knowing (an arbitrary) $\Theta
_{+}(\lambda )$ the previous expression defines $\Theta (\lambda ),$ which
in turn through Eq. (\ref{legendre}) provides the LDF $\varphi (\mu _{\tau
}).$ By construction, the fulfillment of conditions (\ref{phi}) and (\ref%
{teta}) is guaranteed. Alternatively, from Eq. (\ref{ZetalesInversa}) we can
also write%
\begin{equation}
\Theta (\lambda )=\Theta _{-}(-\lambda )+\Theta _{-}(\lambda -\alpha
)-\Theta _{-}(-\alpha ),  \label{TetaTotalInversa}
\end{equation}%
where now $\Theta _{-}(\lambda )$ is the LDF associated to the generating
function of $\mu _{\tau }^{-}.$ The relations (\ref{TetaTotal}) and (\ref%
{TetaTotalInversa}) define the second main result of this paper. They allow
to characterizing the long time regime in terms of asymptotic properties of
the independent stochastic contributions.

\section{Intermittent variables}

Our results allows to build up a variable that satisfy the FR Eq. (\ref%
{FTTemporal}) after knowing the statistical properties of an arbitrary one.
The regime of interest is the stationary one, where the FR symmetry is
characterized by Eq. (\ref{TetaTotal}) or (\ref{TetaTotalInversa}). For
example one can assume Gaussian or Poissonian statistics. In this last case,
if one take $\Theta _{+}(\lambda )=\gamma (1-e^{-\lambda }),$ which
correspond to the LDF of an unidirectional Poisson process with rate $\gamma
,$ from Eq. (\ref{TetaTotal}) we get $\Theta (\lambda )=\gamma
(1-e^{-\lambda })+\gamma e^{-\alpha }(1-e^{\lambda }).$ We note that this
expression corresponds to the LDF for the entropy production of an
asymmetric random walk \cite{UDOColloidal}. Hence, while the previous
analysis seem to be rather abstract, there exist non-trivial dynamics where
they apply.

The random walk model is able to fit some non-Gaussian properties found in
Ref. \cite{UDOColloidal}. Here, motivated by the experimental results of
Ref. \cite{soodPituto}, we introduce a similar generalized model able to
develop intermittence. On the basis of previous analysis, the velocity [see
Eq. (\ref{mumu})] is defined as $v_{st}(t)=v_{st}^{+}(t)-v_{st}^{-}(t),$
where%
\begin{equation}
v_{st}^{\pm }(t)=x_{0}\sum\nolimits_{i}\delta (t-t_{i}^{\pm }).
\end{equation}%
The constant $x_{0}$ introduces the right units of $v_{st}(t),$ $\delta (t)$
is the Dirac delta function, and $t_{i}^{\pm }$ are successive random times.
Consistently, $x_{st}(t)=x_{st}^{+}(t)-x_{st}^{-}(t),$%
\begin{equation}
x_{st}^{\pm }(t)=x_{0}n_{st}^{\pm }(t),  \label{equis}
\end{equation}%
where $n_{st}^{\pm }(t)$\ is the number of ($\pm $delta) events in the
interval $(0,t).$ Hence, they are positive (discrete) random variables.
Their generating functions are denoted as $Z_{n}^{\pm }(s,t)=\langle
e^{-sn_{st}^{\pm }(t)}\rangle =\sum_{m=0}^{\infty }q_{m}^{\pm }(t)e^{-sm},$
where $\{q_{m}^{\pm }(t)\}_{m=0}^{\infty }$ are the respective counting
probabilities. Independently of their statistics, by choosing $n_{st}^{+}(t)$
as the \textquotedblleft free\textquotedblright\ variable, this model
satisfy the FR symmetry (\ref{FTTemporal}) after demanding [see Eq. (\ref%
{measures})] the change of measures%
\begin{equation}
q_{m}^{-}(t)=q_{m}^{+}(t)\frac{e^{-s_{0}m}}{\langle
e^{-s_{0}n_{st}^{+}(t)}\rangle },
\end{equation}%
where $\langle e^{-s_{0}n_{st}^{+}(t)}\rangle =Z_{n}^{+}(s_{0},t),$ and $%
s_{0}$ is an arbitrary positive dimensionless constant. Hence, $%
n_{st}^{-}(t) $ can be read as the \textquotedblleft $s_{0}-$%
ensemble\textquotedblright\ associated to $n_{st}^{+}(t)$ \cite{Counting}.
After a change of variables based on Eqs. (\ref{mumu}) and (\ref{equis}),
the constant $\alpha $ reads%
\begin{equation}
\alpha =s_{0}\delta I\equiv s_{0}(I^{+}-I^{-}),  \label{alfa}
\end{equation}%
where $I^{\pm }\equiv -\lim_{t\rightarrow \infty }t^{-1}(\partial /\partial
s)Z_{n}^{\pm }(s,t)|_{s=0},$ or equivalently $\lim_{t\rightarrow \infty
}\langle n_{st}^{\pm }(t)\rangle \simeq I^{\pm }t,$ which in turn implies $%
\left\langle v\right\rangle =x_{0}(I^{+}-I^{-}).$

In order to close the model, it is necessary to specify the statistical
properties of $n_{st}^{+}(t).$ Its generating function is written as%
\begin{equation}
Z_{n}^{+}(s,t)=Z_{A}^{+}(s,t)+Z_{I}^{+}(s,t),  \label{ZPlus}
\end{equation}%
where the evolution of each contribution read 
\begin{eqnarray}
\frac{dZ_{A}^{+}(s,t)}{dt} &=&-\theta _{s}Z_{A}^{+}(s,t)-\Gamma
_{A}Z_{A}^{+}(s,t)+\Gamma _{I}Z_{I}^{+}(s,t),  \notag \\
\frac{dZ_{I}^{+}(s,t)}{dt} &=&+\Gamma _{A}Z_{A}^{+}(s,t)-\Gamma
_{I}Z_{I}^{+}(s,t),  \label{PoissonModulado}
\end{eqnarray}%
with $\theta _{s}\equiv \gamma (1-e^{-s}).$ These dynamics allow us to read $%
n_{st}^{+}(t)$ as a modulated Poissonian (counting) process \cite{Counting},
whose rate at random times adopts the values $\gamma _{A}=\gamma $ (Active
regime) and $\gamma _{I}=0$ (Inactive regime). The switching between both
states is governed by a classical master equation with transition rates $%
\Gamma _{A}$ and $\Gamma _{I}.$ The asymmetric random walk model \cite%
{UDOColloidal}\ is recovered from Eq. (\ref{PoissonModulado}) by taking $%
\Gamma _{A/I}=0.$

The generating function $Z_{n}^{-}(s,t)$ can be obtained from Eq. (\ref%
{Zetales}) after knowing $Z_{n}^{+}(s,t).$ Hence, the (lattice) distribution 
\cite{vanKampen} of the process $x_{st}(t)$ can be obtained by finding $%
Z_{n}^{+}(s,t)$ from Eq. (\ref{PoissonModulado}) and a posterior numerical
Fourier inversion of $Z_{n}(s,t)=Z_{n}^{+}(s,t)Z_{n}^{-}(-s,t)$ in the $%
s=-ik $ variable. By using the Markovian property of Eq. (\ref%
{PoissonModulado}), the probability of $x_{\tau }^{\infty }\equiv
\lim_{t\rightarrow \infty }x_{st}(t)$ follows from that of$\ x_{st}(t)$ by
taking stationary initial conditions for the rate fluctuations. The
distribution $P(\mu _{\tau })$ of $\mu _{\tau }$ follows from the change of
variables defined by Eq. (\ref{mumu}). 
\begin{figure}[tbp]
\includegraphics[bb=40 838 710 1120,angle=0,width=9cm]{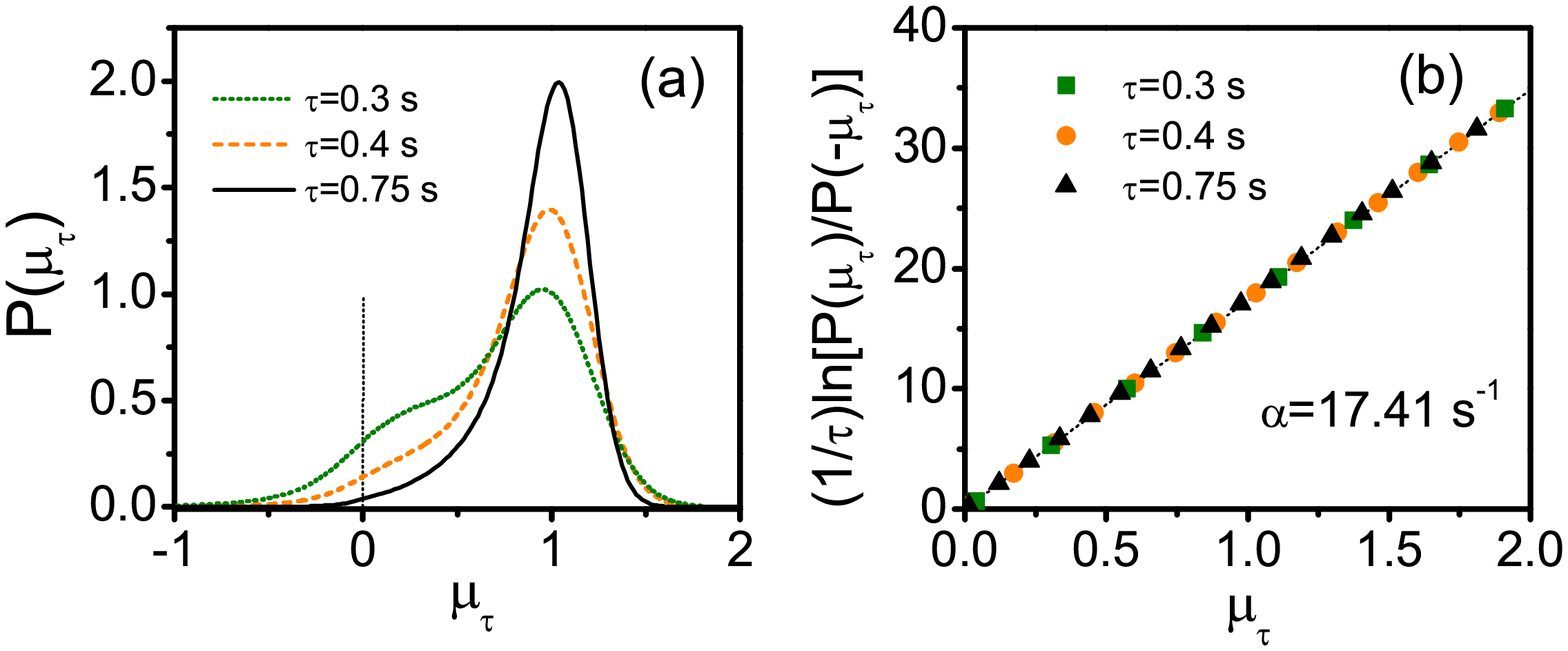}
\caption{(color online) (a) Probability distribution $P(\protect\mu _{%
\protect\tau })$ that follows from Eq. (\protect\ref{PoissonModulado}) (see
text) for different values of $\protect\tau .$ (b) Linear dependence of $(1/%
\protect\tau )\ln \left[ P(+\protect\mu _{\protect\tau })/P(-\protect\mu _{%
\protect\tau })\right] $ corresponding to the distributions shown in (a).
The value of the slope $\protect\alpha $ was obtained from Eq. (\protect\ref%
{alfa}). The parameters [Eq. (\protect\ref{PoissonModulado})] are $%
s_{0}=0.2, $ $\protect\gamma =100s^{-1},\ \Gamma _{A}=0.54s^{-1},$ and $%
\Gamma _{I}=4s^{-1}.$}
\end{figure}

In Fig. 1(a) we plot the distribution $P(\mu _{\tau })$ for different values
of $\tau .$ For all times the distributions satisfy the symmetry Eq. (\ref%
{FTTemporal}) [Fig. 1(b)]. As expected, we confirmed that this property is
valid for any value of the rate parameters that define the evolution (\ref%
{PoissonModulado}). On the other hand, the short time behavior of $P(\mu
_{\tau })$ strongly depends on the chosen parameters values. Nevertheless,
for increasing $\tau $ the distributions becomes similar and, consistently,
develop an increasing peak around $\mu _{\tau }=1.$ In this regime [$\tau
\gtrsim 0.3$ in Fig. 1(a)] a large deviation theory applies. Hence the
problem can be analyzed through the corresponding LDF.

\subsection*{Large deviations functions}

The long time behavior of $Z_{n}^{+}(s,t)$ completely define the asymptotic
statistical properties of the counting process $n_{st}^{+}(t).$ Before
characterizing the LDF functions from it, we notice that in the long time
limit some results can be established for $Z_{n}^{-}(s,t).$ As demonstrated
in Ref. \cite{Counting}, $n_{st}^{+}(t)$ can be mapped with a renewal
process characterized by a shift closure property, which implies that
asymptotically $n_{st}^{-}(t)$ also becomes a renewal process with a
renormalized waiting time distribution. Hence, asymptotically $%
Z_{n}^{-}(s,t) $ has the same structure and dynamics than $Z_{n}^{+}(s,t)$
but with renormalized rates. After some hard calculations steps, which are
not relevant for the following analysis, we obtained $\gamma \rightarrow
\gamma _{s_{0}}=\gamma e^{-s_{0}},$ and more complex expressions for the
hopping rates $\Gamma _{A/I}\rightarrow \Gamma _{A/I}(s_{0}).$ Furthermore $%
I^{+}=\gamma \Gamma _{I}/(\Gamma _{A}+\Gamma _{I}),$ and similarly $%
I^{-}=\gamma _{s_{0}}\Gamma _{I}(s_{0})/[\Gamma _{A}(s_{0})+\Gamma
_{I}(s_{0})].$ In the case of an asymmetric random walk model \cite%
{UDOColloidal},\ $\Gamma _{A/I}=0,$ both set of probabilities $\{q_{m}^{\pm
}(t)\}_{m=0}^{\infty }$ become Poissonian counting processes with $%
I^{+}=\gamma ,$ and $I^{-}=\gamma e^{-s_{0}}.$

By working Eqs. (\ref{ZPlus}) and (\ref{PoissonModulado}) in a Laplace
domain, $f(\xi )=\int_{0}^{\infty }f(t)e^{-\xi t},$ it is possible to write $%
Z_{n}^{+}(s,t)$ as a superposition of two exponential functions scaled by
the roots $[Q(\xi )=0]$ of the characteristic polynomial 
\begin{equation}
Q(\xi )=\xi ^{2}+\xi (\theta _{s}+\Gamma _{A}+\Gamma _{I})+\theta _{s}\Gamma
_{I}.  \label{Polinomio}
\end{equation}%
From its definition, the smaller solution \cite{UDOColloidal}, after the
change of variable $s\rightarrow \lambda /\delta I,$ provides $\Theta
_{+}(\lambda ).$ We get%
\begin{equation}
\Theta _{+}(\lambda )=\frac{\theta _{\lambda }^{\prime }+\Gamma _{A}+\Gamma
_{I}}{2}-\Big{[}\Gamma _{A}\theta _{\lambda }^{\prime }+\frac{1}{4}(\Gamma
_{A}+\Gamma _{I}-\theta _{\lambda }^{\prime })^{2}\Big{]}^{1/2},
\label{tetaPlus}
\end{equation}%
where $\theta _{\lambda }^{\prime }\equiv \theta _{\lambda /\delta I}=\gamma
(1-e^{-\lambda /\delta I}).$ This result, joint with Eq. (\ref{TetaTotal})
and the transformation (\ref{legendre}) completely characterize the
asymptotic statistics and LDF of $\mu _{\tau }.$ 
\begin{figure}[tbp]
\includegraphics[bb=55 560 715 1110,angle=0,width=9cm]{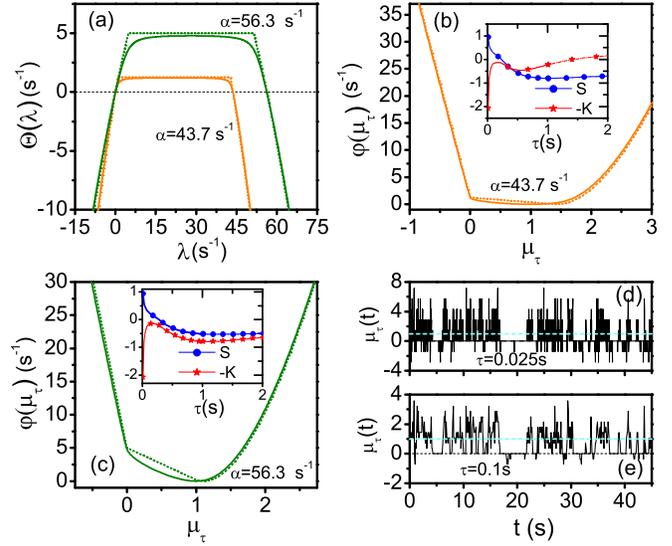}
\caption{(color online) (a) LDF $\Theta (\protect\lambda )$ [Eq. (\protect
\ref{TetaTotal})] for two different values of $\protect\alpha .$ (b)-(c) LDF 
$\protect\varphi (\protect\mu _{\protect\tau })$ [Eq. (\protect\ref{legendre}%
)] for $\protect\alpha =43.7s^{-1}$ and $\protect\alpha =65.5s^{-1}$
respectively. (d)-(e) Realizations of $\protect\mu _{\protect\tau }(t)$ $[%
\protect\alpha =43.7s^{-1}]$ for $\protect\tau =0.025s$ and $\protect\tau %
=0.1s$ respectively. In all cases, the dotted lines follows from
approximation (\protect\ref{Aprox}). The insets show the skewness (S) and
kurtosis (K) as function of $\protect\tau $ indicating the strongly
non-Gaussian nature of $\protect\mu _{\protect\tau }.$ The parameters [Eq. (%
\protect\ref{PoissonModulado})] for the curves with $\protect\alpha %
=43.7s^{-1}$ are $s_{0}=1.56,$ $\protect\gamma =40s^{-1},$ $\Gamma
_{A}=0.54s^{-1},$ and $\Gamma _{I}=1.25s^{-1}.$ For $\protect\alpha %
=56.3s^{-1}$ are the same except $\Gamma _{I}=5s^{-1}.$}
\end{figure}

In Fig. 2(a) we plot the LDF $\Theta (\lambda )$ for two different set of
parameter values $\{s_{0},\gamma ,\Gamma _{A},\Gamma _{I}\}.$ In both cases
the symmetry (\ref{teta}) is satisfied. The values of $\alpha $ were
determined from (\ref{alfa}). Notice that $\Theta (\lambda )$ cannot be well
approximated by a quadratic polynomial, which in turn implies the presence
of strong non-Gaussian features. Figs. 2(b) and 2(c) confirm this fact. Both
LDF $\varphi (\mu _{\tau })$\ were obtained numerically through the Legendre
transformation (\ref{legendre}). The same results follows from the
asymptotic behavior of the exact distribution $P(\mu _{\tau })$ obtained by
Fourier inversion of its generating function. Consistently, the LDF and the
asymptotic exact distribution satisfy the FR (\ref{phi}). In the insets, we
plot the time dependence of the skewness $S=\left\langle \delta \mu _{\tau
}^{3}\right\rangle /\sigma ^{3}$ and kurtosis $K=\left\langle \delta \mu
_{\tau }^{4}\right\rangle /\sigma ^{4}-3,$ obtained from the generating
function of $P(\mu _{\tau }),$ where $\delta \mu _{\tau }=\mu _{\tau
}-\left\langle \mu _{\tau }\right\rangle $ and $\sigma ^{2}=\left\langle
\delta \mu _{\tau }^{2}\right\rangle .$ These objects also confirm the
strong non-Gaussian nature of $P(\mu _{\tau }).$

The functions $\varphi (\mu _{\tau })$ not only depart from a quadratic
polynomial, but also develop a \textquotedblleft kink,\textquotedblright\
that is, and abrupt change around the origin. This characteristic arises
when $\Theta (\lambda )$ presents a plateau regime centered around $\alpha
/2.$ For the chosen parameters values, the curves and these special features
are similar \cite{figuras} to those found in the numerical and experimental
results of Refs. \cite{UDOColloidal,soodPituto}. In the present model, a
physical effect can be associated to these properties: the kink and the
plateau regime are closely related with the development of intermittence in
the stochastic realizations of $\mu _{\tau }.$ This is the third main result
of this contribution.

The stochastic realizations of $n_{st}^{+}(t)$ can be obtained from Eq. (\ref%
{PoissonModulado}) by using standard monte Carlo methods. Alternatively, a
more simple algorithm follows by mapping Eq. (\ref{PoissonModulado}) with a
renewal process \cite{Counting}. On the other hand, the realizations of $%
n_{st}^{-}(t)$ can be obtained from a conditional scheme defined in Ref. 
\cite{Counting}, where realizations of $n_{st}^{+}(t)$ with $m$-events are
selected as one of $n_{st}^{-}(t)$ with probability $e^{-s_{0}m}$ and
discarded with probability $(1-e^{-s_{0}m}).$ Statistically independent
realizations of $n_{st}^{\pm }(t)$ allow to generate the stochastic
trajectories of $\mu _{\tau }$ [see Eqs. (\ref{mumu}) and (\ref{equis})]. 
\begin{figure}[tbp]
\includegraphics[bb=55 560 715 1110,angle=0,width=9cm]{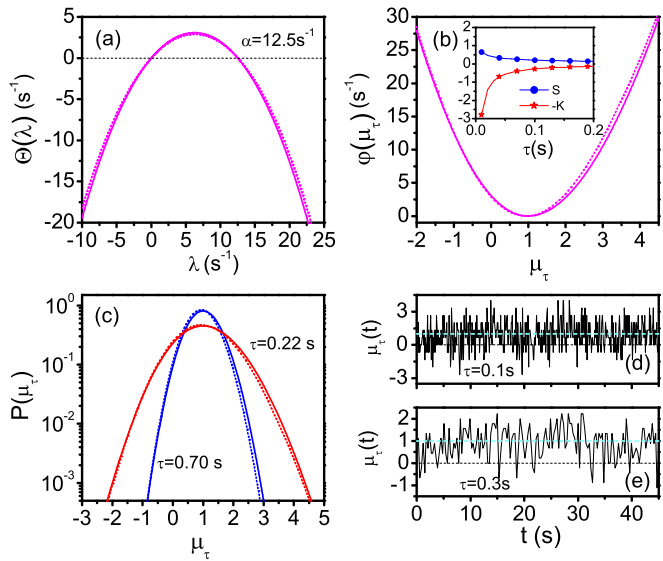}
\caption{(color online) (a) LDF $\Theta (\protect\lambda )$ [Eq. (\protect
\ref{TetaTotal})] with $\protect\alpha =12.5s^{-1}.$ (b) LDF $\protect%
\varphi (\protect\mu _{\protect\tau })$ [Eq. (\protect\ref{legendre})]. (c)
Probability distribution of $\protect\mu _{\protect\tau }$ for $\protect\tau %
=0.22s$ and $\protect\tau =0.70s.$ (d)-(e) Realizations of $\protect\mu _{%
\protect\tau }(t)$ for $\protect\tau =0.1s$ and $\protect\tau =0.3s$
respectively. Inset show the skewness (S) and kurtosis (K) as function of $%
\protect\tau $ indicating tendency to become more Gaussian at higher time $%
\protect\tau .$ In all curves the parameters [Eq. (\protect\ref%
{PoissonModulado})] are $s_{0}=0.83,$ $\protect\gamma =25s^{-1},$ $\Gamma
_{A}=0.3s^{-1},$ $\Gamma _{I}=20s^{-1}.$ The dotted lines follows from
approximation (\protect\ref{Aprox}).}
\end{figure}

In Figs. 2(d) and (e) we plot two realizations of $\mu _{\tau },$
corresponding to $\alpha =43.7s^{-1},$ for two different values of $\tau .$
Consistently with the definition (\ref{mumu}) they fluctuates around $\mu
_{\tau }=1.$ For the chosen parameter values, the trajectories switch
amongst periods of time where $\mu _{\tau }$ is positive, negative or null.
As expected, for increasing $\tau $ the fluctuations are diminished.
Furthermore, for higher averaging times (not shown) the realizations only
assume positive values. From Eq. (\ref{PoissonModulado}), it is immediate to
deduce that when $\gamma \gg (\Gamma _{A},\Gamma _{I}),$ the realizations of 
$n_{st}^{+}(t)$ develop intermittence. This is not a surprising result. What
is novel and not trivial, is that one can build up the complementary process 
$n_{st}^{-}(t)$ such that their (normalized) subtraction [Eq. (\ref{mumu})]
satisfy the FR. The closeness between the kinks of Figs. 2(b) and (c) and
the intermittence shown in Figs. 2(d) and (e) is also supported by the
results shown in Fig. 3. Here, $\Gamma _{I}\gg \Gamma _{A}.$ Therefore, the
inactive regime is statistically (dynamically) inhibited, which in turn
implies that neither $n_{st}^{+}(t)$ or $n_{st}^{-}(t)$ develops any
intermittence. As shown by the plots, the LDF converge to quadratic
polynomial functions, while the probability distribution becomes Gaussian.
Consistently, the trajectories, Figs. 3(d) and (e), for any value of $\tau $
do not develop the phenomenon of intermittence.

Eqs. (\ref{TetaTotal}) and (\ref{tetaPlus}) provide an analytical expression
for the LDF $\Theta (\lambda )$ that is very complicated. Furthermore, the
transformation Eq. (\ref{legendre}), which delivers the LDF $\varphi (\mu
_{\tau }),$ is only manageable numerically. Nevertheless, when $\Gamma
_{A}\ll (\gamma ,\Gamma _{I}),$ from Eq. (\ref{Polinomio}) it is simple to
deduce the expression%
\begin{equation}
\lim_{\Gamma _{A}\rightarrow 0}\Theta _{+}(\lambda )=\min [\theta _{\lambda
}^{\prime },\Gamma _{I}].  \label{Aprox}
\end{equation}%
This result, joint with Eqs. (\ref{legendre}) and (\ref{TetaTotal}) allow us
to obtain simple analytical expressions for the LDF $\Theta (\lambda )$ and $%
\varphi (\mu _{\tau })$ (see Appendix). They correspond to the dotted lines
of Figs. 2 and 3. Evidently, they provide a very good fitting to the plotted
curves. Nevertheless, what is more relevant is the possibility of getting an
analytical description of the relation between intermittence and the
non-Gaussian features of both LDF.

In the limit $\Gamma _{A}\rightarrow 0,$ both $\Theta (\lambda )$ and $%
\varphi (\mu _{\tau })$ present points where they are not derivable
functions with respect to their arguments. In particular, when intermittence
develops, $\Gamma _{I}\ll \gamma ,$ the LDF $\varphi (\mu _{\tau })$ has a
linear behavior around the origin with different slopes for $\mu _{\tau
}\lessgtr 0$ [see dotted lines $(\mu _{\tau }\gtrless 0)$\ in Figs. 2(b) an
(c)]. The addition of the slopes of $\varphi (\mu _{\tau })$ around the
origin is equal to $-\alpha $ [see Eq. (\ref{phiDesnudaInter})]. The
experimental data of Ref. \cite{soodPituto} seem to be consistent with this
relation. On the other hand, in the intermittence regime $\Theta (\lambda )$
develops a plateau regime [see dotted lines\ in Figs. 2(a)] with value $%
\Gamma _{I}$ [Eq. (\ref{TetolDiscontinuo})], which is similar to that found
in the numerical results of Ref. \cite{UDOColloidal}.

Finally, let us remark that the discontinuities of the derivative of $\Theta
(\lambda )$ can be read as phase transitions between different dynamical
regimes of $\mu _{\tau }$ \cite{garrahan,Counting}. These features, as well
as our main proposal, Eq. (\ref{measures}), suggest an interesting bridge
between FR and equilibrium thermodynamics.

\section{Conclusions}

In conclusion, we have demonstrated that FR are always satisfied by the
subtraction of two independent stochastic variables whose probability
distributions are related by a thermodynamic-like change of measure. This
relation leaves completely arbitrary one of the distributions, being this
degree of freedom the basis of the present approach. By choosing one of then
as a random modulated Poisson process, the distribution of interest develops
strong non-Gaussian features, which in turn arise in the parameter regime
where intermittence develops. Hence, variables such as entropy production
rates may satisfy FR and develop intermittence.

Depending of the parameters values of the model the LDF develop a rich
variety of functional dependences, which in turn are similar to those found
in recent numerical and experimental results \cite{figuras}. In particular,
in the intermittence regime the LDF of the probability distribution and
generating function are characterized by a kink at the origin and a plateau
regime respectively. While our analysis does not relies on any specific
(non-linear many body) dynamical model, it strongly suggest that
intermittence may be a central ingredient in nonequilibrium steady states
characterized by non-Gaussian fluctuations.

Due to the wide class of systems where intermittence develops, added to the
central role of FR in non-equilibrium states, the characterization of
dynamical conditions that guaranty the coexistence of these two properties
becomes a very interesting open issue, for which this paper intends to
contribute. On other hand, the inclusion of statistical correlations that
preserve the FR in the two-variables model as well as the analysis of these
ideas in the context of deterministic thermostated systems (phase space
contraction conjecture) \cite{searles} are additional open issues that may
deserve extra analysis.

\section*{Acknowledgments}

This work was supported by CONICET, Argentina, PIP 11420090100211.

\appendix*

\section{LDF in the limit $\Gamma _{A}\rightarrow 0$}

Here we obtain the LDF associated to Eq. (\ref{Aprox}). $\Theta (\lambda )$
follows from Eq. (\ref{TetaTotal}). Hence, $\varphi (\mu _{\tau })$ is
obtained through the Legendre transformation (\ref{legendre}). In order to
simplify the expressions, we write 
\begin{equation}
\Theta (\lambda )=\bar{\Theta}(s=\lambda /\delta I),
\end{equation}%
and similarly $\varphi (\mu _{\tau })$ as%
\begin{equation}
\varphi (\mu _{\tau })=\bar{\varphi}(k=\mu _{\tau }\delta I),
\label{phiReescaleo}
\end{equation}%
where $\delta I=s_{0}(I^{+}-I^{-})$ [Eq. (\ref{alfa})]. The functions $\bar{%
\Theta}(s),$\ and $\bar{\varphi}(k)$ are the LDF of the stochastic variable $%
k_{st}(t)=n_{st}^{+}(t)-n_{st}^{-}(t)]/t.$ Furthermore, we define the
parameters%
\begin{eqnarray}
s^{+} &\equiv &-\ln \Big{(}1-\frac{\Gamma _{I}}{\gamma }\Big{)}, \\
s^{-} &\equiv &s_{0}+\ln \Big{(}1-\frac{\Gamma _{I}}{\gamma }\Big{)},
\end{eqnarray}%
and respectively%
\begin{eqnarray}
k^{+} &\equiv &\gamma e^{-s^{+}}=\gamma \Big{(}1-\frac{\Gamma _{I}}{\gamma }%
\Big{)}, \\
k^{-} &\equiv &\gamma e^{-s^{-}}=\gamma \frac{e^{-s_{0}}}{1-\frac{\Gamma _{I}%
}{\gamma }}.
\end{eqnarray}%
The LDF of a Poisson process with rate $\gamma $ is denoted as%
\begin{equation}
\theta (s)\equiv \gamma (1-e^{-s}),
\end{equation}%
and its (symmetrized) Legendre transform as%
\begin{equation}
\varphi _{p}(k)\equiv \gamma \Big{\{}1-\frac{|k|}{\gamma }\Big{[}1-\ln %
\Big{(}\frac{|k|}{\gamma }\Big{)}\Big{]}\Big{\}}.
\end{equation}%
Finally, we introduce the function%
\begin{eqnarray}
\varphi _{rw}(k) &\equiv &\gamma \Big{\{}1+e^{-s_{0}}-\Big{(}4e^{-s_{0}}+%
\frac{k^{2}}{\gamma ^{2}}\Big{)}^{1/2}  \label{phiRW} \\
&&+\frac{k}{\gamma }\ln \Big{[}\frac{1}{2}\Big{(}4e^{-s_{0}}+\frac{k^{2}}{%
\gamma ^{2}}\Big{)}^{1/2}+\frac{k}{2\gamma }\Big{]}\Big{\}}.  \notag
\end{eqnarray}%
Both $\bar{\Theta}(s)$\ and $\bar{\varphi}(k)$ must be defined in different
parameter regimes.

i) In the parameter regime%
\begin{equation}
0<\frac{\Gamma _{I}}{\gamma }<(1-e^{-s_{0}/2}),
\end{equation}%
it follows $s^{+}<s^{-},$ with%
\begin{equation}
\bar{\Theta}(s)=\left\{ \!%
\begin{array}{ccc}
\theta (s), &  & s<s^{+}, \\ 
\Gamma _{I}, &  & s^{+}<s<s^{-}, \\ 
\theta (-s+s_{0}), &  & s^{-}<s,%
\end{array}%
\right.   \label{TetolDiscontinuo}
\end{equation}%
and respectively%
\begin{equation}
\bar{\varphi}(k)=\left\{ \!%
\begin{array}{ccc}
\varphi _{p}(k)-ks_{0}, &  & k<-k^{+}, \\ 
\Gamma _{I}-ks^{-}, &  & -k^{+}<k<0, \\ 
\Gamma _{I}-ks^{+}, &  & 0<k<k^{+}, \\ 
\varphi _{p}(k), &  & k^{+}<k.%
\end{array}%
\right.   \label{phiDesnudaInter}
\end{equation}%
This is one of the more interesting parameter regimes, which in fact
corresponds to the intermittence one (Fig. 2). Notice that in the plateau
regime of $\Theta (\lambda )$ $(s^{+}\delta I<\lambda <\delta Is^{-})$ it
assumes the value $\Gamma _{I}.$ On the other hand, around the origin $%
\varphi (\mu _{\tau })$ has a linear behavior which different slopes for $%
\mu _{\tau }\gtrless 0.$ This property gives rise to the characteristic kink
shown in Figs. 2(b) and (c). From Eqs. (\ref{phiReescaleo}) and (\ref%
{phiDesnudaInter}), we deduce that the addition of the slopes is equal to $%
-\alpha $ [Eq. (\ref{alfa})]. Furthermore, from Eq. (\ref{phiDesnudaInter}),
it is possible to demonstrate that the origin is the unique point at which
the derivative of $\varphi (\mu _{\tau })$ is a discontinuous function.
Remarkably, in the following parameter regimes $\varphi (\mu _{\tau })$ has
a continuous derivative. Hence, a kink related to a discontinuous derivative
only arises in the present case.

ii) In the parameter regime%
\begin{equation}
(1-e^{-s_{0}/2})<\frac{\Gamma _{I}}{\gamma }<(1-e^{-s_{0}}),
\end{equation}%
it follows $s^{-}<s^{+},$ with%
\begin{equation}
\bar{\Theta}(s)=\left\{ \!%
\begin{array}{ccc}
\theta (s), &  & s<s^{-}, \\ 
\theta (s)+\theta (-s+s_{0})-\Gamma _{I}, &  & s^{-}<s<s^{+}, \\ 
\theta (-s+s_{0}), &  & s^{+}<s,%
\end{array}%
\right. 
\end{equation}%
and by defining the parameters%
\begin{equation}
k^{\ast }\equiv k^{-}-k^{+},\ \ \ \ \ \ \ \ \ \Delta \equiv \Gamma
_{I}-\theta (s_{0}),
\end{equation}%
$(k^{\ast }>0),$ we write%
\begin{equation}
\bar{\varphi}(k)=\left\{ \!%
\begin{array}{ccc}
\varphi _{p}(k)-ks_{0}, &  & k<-k^{-}, \\ 
(\gamma -k^{-})-ks^{+}, &  & -k^{-}<k<-k^{\ast }, \\ 
\varphi _{rw}(k)-\Delta , &  & -k^{\ast }<k<k^{\ast }, \\ 
(\gamma -k^{-})-k(s_{0}-s^{+}), &  & k^{\ast }<k<k^{-}, \\ 
\varphi _{p}(k), &  & k^{-}<k.%
\end{array}%
\right. 
\end{equation}%
In this intermediate regime, the non-Gaussian properties are gradually lost.
The plateau regime of $\Theta (\lambda )$ becomes bend. On the other hand,
we remark that even when $\varphi (\mu _{\tau })$ is defined by parts, its
first derivative is a continuous function.

iii) In the parameter regime%
\begin{equation}
(1-e^{-s_{0}})<\frac{\Gamma _{I}}{\gamma }<1,
\end{equation}%
with $s^{-}<s^{+},$ we get%
\begin{equation}
\bar{\Theta}(s)=\left\{ \!%
\begin{array}{ccc}
\theta (s)+\Delta , &  & s<s^{-}, \\ 
\theta (s)+\theta (-s+s_{0})-\theta (s_{0}), &  & s^{-}<s<s^{+}, \\ 
\theta (-s+s_{0})+\Delta , &  & s^{+}<s,%
\end{array}%
\right. 
\end{equation}%
and%
\begin{equation}
\bar{\varphi}(k)\!=\!\left\{ \!%
\begin{array}{cc}
\varphi _{p}(k)-ks_{0}+\Delta , & k<-k^{-}, \\ 
(\gamma -k^{-})-ks^{+}+\Delta , & -k^{-}<k<-k^{\ast }, \\ 
\varphi _{rw}(k), & -k^{\ast }<k<k^{\ast }, \\ 
(\gamma -k^{-})-k(s_{0}-s^{+})+\Delta , & k^{\ast }<k<k^{-}, \\ 
\varphi _{p}(k)+\Delta , & k^{-}<k.%
\end{array}%
\right. 
\end{equation}%
Fig. 3 falls in this regime, where both LDF approach quadratic functions. In
fact, while the derivative of $\Theta (\lambda )$ presents smooth
discontinuities they are far beyond of the origin $[\lim_{(\Gamma
_{I}/\gamma )\rightarrow 1}s^{\pm }=\pm \infty ].$ As in the previous case, $%
\varphi (\mu _{\tau })$ has a continuous derivative. Its dependence is
mainly defined by the function $\varphi _{rw}(k)$ $[\lim_{(\Gamma
_{I}/\gamma )\rightarrow 1}k^{\ast }=\infty ].$

iv) In the parameter regime%
\begin{equation}
1<\frac{\Gamma _{I}}{\gamma }<\infty ,
\end{equation}%
for any value of $s,$ we get%
\begin{equation}
\bar{\Theta}(s)=\theta (s)+\theta (-s+s_{0})-\theta (s_{0}).
\end{equation}%
Therefore, in this approximation $\Theta (\lambda )$ correspond to the LDF
of an asymmetric random walk. Furthermore, it follows $\bar{\varphi}%
(k)=\varphi _{rw}(k)$ [Eq. (\ref{phiRW})]. As in the previous regime, the
rate fluctuations do not induce any non-Gaussian feature because their
characteristic time is comparable to that of the individual events.


\begin{thebibliography}{99}
\bibitem{figuras} Notice that Figs. 2(b), 2(c), and 3(c) are respectively
very similar to Figs. 2(c), 2(d) and 3(b) of Ref. \cite{soodPituto}. In
addition, Figs. 1 of Ref. \cite{UDOColloidal} can be recovered with other
parameter values. Besides these similitudes, it is important to notice that
our analysis only concentrates on asymptotic statistical properties of $\mu
_{\tau }.$ Hence, there may exist different processes $v(t),$ and $x(t),$
that lead to the same long-time statistical behaviors. On the other hand,
the description of specific non-stationary properties, where in general the
FR is not satisfied, as well as the relation between the parameters of the
model and experimental variables, are interesting open problems whose
solutions depend on each specific situation. These issues are beyond the
scope of this contribution.
\end{thebibliography}
\end{document}